\documentclass[twocolumn]{jpsj2} %% two-column layout
\usepackage{txfonts}

\def\runauthor{
\textsc{Kuramoto}, 
\textsc{Kiss}, \textsc{Otsuki}and 
\textsc{Kusunose}
} 
\title{
Hybridization and multipole orders of $4f$ electrons in Pr skutterudites
}

\author{
Yoshio \textsc{Kuramoto}\thanks{E-mail: kuramoto@cmpt.phys.tohoku.ac.jp},
Annam\'{a}ria \textsc{Kiss}, Junya \textsc{Otsuki}  and 
Hiroaki \textsc{Kusunose}
}

\inst{
Department of Physics, Tohoku University, Sendai 980-8578 
}

\abst{
Characteristics of hybridization and multipole orders of $4f$-electrons in Pr skutterudites are explained in terms of a pseudo-quartet composed of crystalline electric field (CEF) singlet and the triplet.  It is shown that the contrasting behaviors observed in 
PrFe$_4$P$_{12}$ and PrOs$_4$Sb$_{12}$ are ascribed to the difference in  triplet wave functions and the CEF splittings.
Since a macroscopic degeneracy remains even in the ordered phase with  
the $\Gamma_3$-type antiferro-quadrupole (AFQ) order,
a model 
with strong AFQ fluctuation and static monopole/hexadecapole order 
is proposed for PrFe$_4$P$_{12}$.
The remaining cubic symmetry explains qualitatively the behavior of staggered magnetization observed by NMR and neutron scattering. 
For identification of possible hexadecapole orders in PrFe$_4$P$_{12}$ and
PrRu$_4$P$_{12}$, 
eightfold intensity pattern is predicted in the azimuthal angle scan of resonant X-ray scattering.   
In PrOs$_4$Sb$_{12}$, the ferromagnetic exchange coupling with the conduction band does not lead to magnetic Kondo effect,
but a momentum-dependent quadrupole coupling can give rise to enhanced effective mass, which should be sensitive to disorder.
}

\kword{
quadrupole order, quantum fluctuation, 
resonant X-ray scattering, skutterudite
}

\begin{document}
\maketitle

\section{Introduction}
There are puzzling properties in filled Pr skutterudite compounds.  
In this paper we concern with mechanism of the following behaviors:
PrFe$_4$P$_{12}$ undergoes a phase transition at 7 K in zero field \cite{PrFeP}. 
The susceptibility shows a cusp at the transition, but there is no N\'{e}el order in the ordered phase.  Then an antiferro-quadrupole (AFQ) order has been considered as a plausible explanation \cite{kiss}.  Indeed a staggered magnetic moment and the lattice distortion have been found by neutron diffraction \cite{hao}.  However, in contrast with the cas of PrOs$_4$Sb$_{12}$ \cite{kohgi}, there is no component perpendicular to the field of the staggered moment \cite{hao,kikuchi}.
If the order parameter is AFQ,
one may naturally expect finite perpendicular component in some directions of the field.
Thus proper identification of the order parameter remains to be done.
On the other hand, PrOs$_4$Sb$_{12}$ has well-defined CEF excitations as seen by neutron scattering \cite{kohgi,gore}, and by the Schottky-type anomaly in the specific heat \cite{bauer,aoki-jpsj}.  Near the Schottky peak around 2 K, superconductivity sets in with a huge discontinuity in the specific heat.  The problem here is the origin of the Cooper pair.  
More modestly, one may ask the mechanism which produces heavy electrons forming the Cooper pair.
These two Pr skutterudites also show contrasting behaviors in the temperature dependence of the resistivity; PrFe$_4$P$_{12}$ shows a typical Kondo effect, while PrOs$_4$Sb$_{12}$ only a mild shoulder in the temperature dependence \cite{sato}.  The inelastic neutron spectrum of PrFe$_4$P$_{12}$ is characterized by broad quasi-elastic features like other heavy-fermion systems, without well defined CEF excitations \cite{Iwasa_Fe}. 

In this paper we propose a possible route toward understanding these fundamental issues.
We begin with brief review of our previous results which explain the contrasting behaviors in terms of the character of CEF states.
Then we proceed to rather speculative new ideas.  These include proposal of a novel multipole order in PrFe$_4$P$_{12}$ in which fluctuating quadrupoles are not frozen even in the ordered phase.
The frozen multipole is a hexadecapole which should be probed by resonant X-ray scattering.
In addition, for PrOs$_4$Sb$_{12}$,
we propose that the induced quadrupolar fluctuation with momentum dependent coupling with conduction band is responsible for the heavy mass. 

\section{Local electronic states and hybridization effects}
\subsection{CEF splitting by p-f hybridization} 
Filled skutterudites RT$_4$X$_{12}$ form a bcc structure with the space group $Im\bar{3}$.
The R site has the local symmetry $T_h$ which has no four-fold rotation axis.\cite{THY}
In this symmetry, the $4f^2$ Hund's-rule ground states $^3H_4$ of Pr$^{3+}$ split into a singlet $\Gamma_1$, a 
non-magnetic doublet $\Gamma_{23}$, and two $\Gamma_4$ triplets.  
Of these, 
two $\Gamma_4$ triplets are written as $\Gamma_4^{(1)}$ and $\Gamma_4^{(2)}$, which are linear combinations of triplets $\Gamma_4$ and $\Gamma_5$ in $O_h$.
Under this crystal symmetry, the CEF potential is written as
\begin{align}
V_{\text{CEF}} 
&	 = A_4 [O_4^0+5O_4^4]+A_6^{\text{c}}[O_6^0-21O_6^4]+A_6^{\text{t}}[O_6^2-O_6^6]
\nonumber \\
& = W\left[ x \frac{O_4}{60} +(1-|x|)\frac{O_6^{\rm c}}{1260}+y\frac{O_6^{\rm t}}{30}
 \right],
\label{V_CEF}
\end{align}
in the standard notation\cite{THY}.
The term $yO_6^{t}$ mixes the $\Gamma_4$ 
and $\Gamma_5$  triplet states in the point group $O_h$.  

\begin{figure}[bt]
\centerline{
\includegraphics[width= 0.5\textwidth]{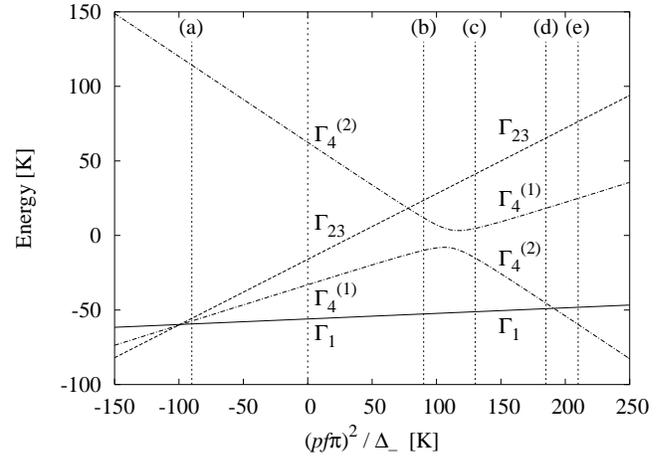}
}
\caption{
CEF level structures derived from hybridization and point charge potential as a function of 
$(pf\pi)^2/\Delta_-$. 
The level sequence qualitatively corresponds to: 
(a) PrFe$_4$P$_{12}$;
(b) PrRu$_4$P$_{12}$ in the high-temperature phase; 
(c) Pr1 site in PrRu$_4$P$_{12}$ in the low-temperature phase; 
(d) PrOs$_4$Sb$_{12}$;
(e) Pr2 site in PrRu$_4$P$_{12}$ in the low-temperature phase.  
See text for details.
}
\label{cef}
\end{figure}

There are two main sources for the CEF splitting: 
the Coulomb potential from ligands, and hybridization between $4f$ electrons and ligands.
In the point charge model, CEF coefficients $A_4, A_6^{\rm c}$ and $A_6^{\rm t}$ in eq.(\ref{V_CEF}) are determined by coordination of the charge and the radial extension of the $4f$ wave function. 
Explicit results have been obtained in ref.\citen{Otsuki1} taking the effective charge $Z_{\rm t}$ of a transition ion as the parameter and
requiring the charge neutrality with trivalent Pr.
To estimate the CEF potential, 
we use the lattice parameter in PrOs$_4$Sb$_{12}$\cite{Sugawara}, which gives
the Pr-T distance $d_{\rm t}=4.03\AA$ with T=Sb, the Pr-X distance $d_{\rm p}=3.48\AA$ with X=Os and the 
X-Pr-X vertical angle $2\theta_0=49.2 
^\circ$.
These data will be used to derive the results presented in Fig.\ref{cef}.

Another important mechanism for CEF splittings is the covalency effect, or hybridization between localized and ligand orbitals.
According to band calculation,\cite{Harima} 
the conduction band striding  the Fermi level is formed mostly by the molecular orbital $a_u$ formed by 12 pnictogens around each Pr. 
The other relevant molecular orbital $t_u$ 
form two energy bands around a few eV above the Fermi level, and three a few eV below. 
We neglect contributions from $t_u$ bands to CEF splittings 
 because of the larger excitation energy. 
Therefore we first consider hybridization of the form
\begin{equation}
H_{\text{hyb}}
=V_{a}\sum_{\sigma}f^{\dag}_{\sigma} c_{\sigma} + \text{h.c.},
\end{equation}
where $c_{\sigma}$ annihilates a conduction electron in the Wannier orbital with the $a_u$ 
symmetry at the origin,
and $f^{\dag}_{\sigma}$ creates an $4f$ electron with the same ($a_u$) orbital symmetry.
We take the hybridization parameter $V_{a}$ real.
Hereafter we adopt the Mulliken notation such as $a_u$ for orbital symmetry, and the Bethe notation such as $\Gamma_1$ for a double-group representation with spin-orbit coupling.
In the second-order perturbation theory, the effective interaction is given by
\begin{equation}
	H_{\rm eff}= P H_{\text{hyb}} \frac{1}{E-H_0} Q H_{\text{hyb}} P,
\label{H_int}
\end{equation}
where $P$ is the projection operator onto 
$4f^2$ states, and $Q=1-P$.
We first deal with such part of the second-order hybridization that is diagonal with respect to the conduction states.
Diagonalization of $H_{\rm eff}$ with this constraint gives the 
CEF wave functions and their energies.  
We assume the following intermediate states:
(i) 4$f^{1}$ and an extra electron in vacant states, and
(ii) 4$f^{3}$ and extra hole in filled states.
For simplicity, we neglect the multiplet splittings in both kinds of intermediate states. 
We note that  $4f^1$ and $4f^3$ intermediate states give opposite contributions to the level splitting.
Hence the sequence of CEF levels is determined by competition between both intermediate states.
The hybridization is parameterized by the Slater-Koster parameters $(pf\sigma)$ and $(pf\pi)$.
It turns out hybridization with $a_u$ comes only from $(pf\pi)$.

We  combine both contributions to CEF splittings by point-charge
interaction and by hybridization.  
To a good approximation we can assume the half-filled $a_u$ band.
We introduce a parameter
$$
1/\Delta_-=1/\Delta_3 -1/\Delta_1.
$$
Note that $1/\Delta_-=0$ in the case of $\Delta_3=\Delta_1$, and $1/\Delta_-$ becomes negative if $4f^1$ is the dominant intermediate state.
For the point charge parameters we tentatively take $Z_{\rm t}=2$.
Figure \ref{cef} shows computed results 
as a function of strength of hybridization, $(pf\pi)^2/\Delta_-$.
It should be emphasized that the the closely lying singlet and triplet are realized
only with simultaneous action of point-charge and hybridization interactions.

The lowest triplet $\Gamma_{\rm t}$, which is  either $\Gamma_4^{(1)}$ or $\Gamma_4^{(2)}$, can be written in terms of $O_h$ triplets $\Gamma_4$ and $\Gamma_5$ as
\begin{eqnarray}
|\Gamma_{\rm t},m\rangle
=\sqrt{w}|\Gamma_4,m\rangle+\sqrt{1-w}|\Gamma_5,m\rangle,
\end{eqnarray}
where $m=\pm, 0$ specifies a component, and $w$ with $0<w<1$ gives the weight of $\Gamma_4$ states.
We have chosen the phase of wave functions so that positive coefficients give the lower triplet.
The level repulsion between $\Gamma_4^{(1)}$ and $\Gamma_4^{(2)}$ around $(pf\pi)^2/\Delta_-=110$K is due to mixing of wave functions, and is a characteristic feature in the point group $T_h$.   
Possible location of effective hybridization in some representative Pr skutterudites is schematically shown by the lines (a)-(e) in Fig.\ref{cef}.  
In the case of PrFe$_4$P$_{12}$, the lattice constant is 7.81$\AA$ as compared with 9.3$\AA$ of PrOs$_4$Sb$_{12}$.
Hence $(pf\pi)$ in PrFe$_4$P$_{12}$ should be larger.
At first sight, the larger hybridization in PrFe$_4$P$_{12}$ seems incompatible with this leftward shift (a) from PrOs$_4$Sb$_{12}$ around (d).
However, since the smaller lattice constant in PrFe$_4$P$_{12}$ favors the $4f^1$ excited states rather  than the $4f^3$ states,  the parameter $\Delta_1$ can become smaller, leading to negative $\Delta_-$.
Since the hybridization with $a_u$ orbitals alone cannot take into account the parameter $y$ in eq.(\ref{V_CEF}),  the simultaneous crossing of three levels near (a) should be taken with caution.  In other words, only the Coulomb potential is the source of the $y$ parameter in our result.
From the study of phase diagram in magnetic field \cite{kiss05},
it seems that the level $\Gamma_{23}$ is situated higher in this region.  As a result,  if high pressure drives the system toward further left in the Fig.\ref{cef}, it is likely that the triplet becomes the ground state instead of the doublet. 

It is reasonable that the larger cage in PrOs$_4$Sb$_{12}$ prefers $4f^3$ intermediate states.
In PrOs$_4$Sb$_{12}$, the triplet forming pseudo-quartet is almost of the $\Gamma_5$-type as shown by (d).  On the other hand, the CEF levels in the high-temperature phase of PrRu$_4$P$_{12}$ corresponds to (b), which bifurcates into (c) and (e) in the low-temperature phase.  
Namely in one sublattice called Pr1, the triplet changes the character slightly, while in the other sublattice called Pr2, the triplet becomes the ground state.  Below the phase transition, the system behaves like an insulator.   It may be possible to describe the transition as the antiferro-hexadecapole order as discussed elsewhere \cite{yukawa,takimoto,hiro}.  

\subsection{Exchange interactions with conduction electrons}

We are interested in the matrix element between the singlet and the triplet in the $T_h$ group.   As we have seen the triplet is a linear combination of $\Gamma_4$ and $\Gamma_5$ in the $O_h$ group.  
Hence it is convenient to analyze the symmetry properties of triplets in the cubic case.  
The direct products of pseudo-quartets can be decomposed as 
\begin{align}
	(\Gamma_1 \oplus \Gamma_4) \otimes (\Gamma_1 \oplus \Gamma_4)
	 &= \Gamma_1 + 2\Gamma_4 + (\Gamma_1 + \Gamma_3 + \Gamma_4 + \Gamma_5), \\
	(\Gamma_1 \oplus \Gamma_5) \otimes (\Gamma_1 \oplus \Gamma_5)
	 &= \Gamma_1 + 2\Gamma_5 + (\Gamma_1 + \Gamma_3 + \Gamma_4 + \Gamma_5),
\label{product}	 
\end{align}
where $\Gamma_4$ has the same symmetry as the magnetic moment. 
The product of 
$\Gamma_1\oplus\Gamma_4$ produces newly two $\Gamma_4$ representations, of which one is time-reversal odd and the other is even.
The odd representation represents the magnetic moment, while the even one the hexadecapole moment.
On the other hand, the pseudo-quartet $\Gamma_1 \oplus \Gamma_5$
 does not produce a new $\Gamma_4$ representation.
Physically, this means that $\Gamma_4$ as the first excited states gives rise to a van Vleck term in the magnetic susceptibility, while $\Gamma_5$ does not.

It is convenient to introduce the effective moment operators in the pseudo-quartet in $T_h$.  In the case of $w=0$ (pure $\Gamma_5$ triplet), 
the magnetic moment within $\Gamma_{\rm t}$ is the only relevant quantity.  The vector operator in this case is written as  $\mib{X}^{\rm t}$.
On the other hand, another vector operator 
$\mib{X}^{\rm s}$ is necessary to describe the magnetic moment of van Vleck type.  
The operator $\mib{X}^{\rm t}$ act on the triplet states, and $\mib{X}^{\rm s}$ connect the singlet and the triplet.
The effective interaction within the pseudo-quartet is given by
\begin{align}
	H_{\text{exc}} &= 
	\left( 
	 I_{\rm t} \mib{X}^{\rm t} + I_{\rm s} \mib{X}^{\rm s}
	  \right)\cdot 
\frac 12\sum_{\alpha\beta}
c^{\dag}_{\alpha} 
\mib{\sigma}_{\alpha \beta}c_{\beta} .
\label{XtXs}	  
\end{align}
where the conduction electron spin 
$\mib{s}_c$ is explicitly written in terms of fermion operators.
The coupling constants are proportional to $V_{a}^2$
and negative \cite{Otsuki1}.
The origin of the ferromagnetic exchange
is traced to the Hund rule involving the spin-orbit interaction; the dominant orbital moment is pointing oppositely from the spin moment.  
In other words, the spin exchange is antiferromagnetic as is usual for a hybridization induced exchange.  

Alternatively, one can use
the pseudo-spin representation as in ref.\citen{shiina-aoki}: 
$$
\mib{X}^{\rm t} = \mib{S}_1 + \mib{S}_2,\ 
\mib{X}^{\rm s} = \mib{S}_1 - \mib{S}_2,
$$
where $\mib{S}_1$ and $\mib{S}_2$ are spin 1/2 operators.
The pseudo-spin representation naturally leads to 
commutation rules among
$\mib{X}^{\rm t}$ and $\mib{X}^{\rm s}$. 
The exchange interaction in terms of pseudo-spins is given by
\begin{equation}
	H_{\text{exc}} = 
	(J_1\mib{S}_1 + J_2\mib{S}_2)\cdot \mib{s}_c,
\label{pseudospin}	
\end{equation}
where 
$J_1= I_{\rm t}+ I_{\rm s}$ and 
$J_2= I_{\rm t}- I_{\rm s}.
$
It turns out 
that $J_2$ becomes positive for as $w$ deviates from 0.
The emergence of antiferromagnetic exchange is due to the particular CEF level structure in Pr skutterudites.   
$J_2$ is almost negligible in the pure $\Gamma_5$ case ($w=0$), and becomes an order of magnitude larger as $w$ increases toward unity, {\it i.e.}, toward pure $\Gamma_4$.  
It is likely that the Kondo-type behavior seen in PrFe$_4$P$_{12}$ originates from a $\Gamma_4$-type triplet
together with a small singlet-triplet splitting.
Nemerical results have been reported for dynamical properties of the pseudo-quartet \cite{Otsuki2} with use of the NCA \cite{nca1}.

\subsection{Quadrupole interactions with conduction electrons}

Although the conduction band is of $a_u$ type, it can acquire the coupling with quadrupole moments through mixing with $t_u$-type calcogen bands.   We start with the 
most general form of  p-f hybridization as given by
\begin{equation}
H_{\text{hyb}} = 
\sum_{\Gamma(\alpha\beta)}\sum_{\nu \sigma} \left[ V_{\Gamma{(\alpha\beta)}}
	 p_{\Gamma{(\alpha)} \nu \sigma}^{\dag} f_{\Gamma{(\beta)} \nu \sigma}
	 +\text{H.c.} \right],
\end{equation}
where the indices $\alpha,\beta$ 
distinguish different sets with the same irreducible representation $\Gamma$.
A member in the irreducible representation is specified by  $\nu$,
and $\sigma$ denotes a spin component. 
Let us consider nine Pr sites making up a body-centered cube each of which is surrounded by 12 pnictogens.  A 4f orbital state $3z^3-5$,
for example, in the body centered Pr at $\mib{R}=(000)$ has hybridization with surrounding 
pnictogen orbitals $\Gamma=t_u, \nu=z$, for which 
five different sets have been enumerated \cite{Otsuki1}.

Each $t_u$ molecular orbital mixes with the eight $a_u$ orbitals surrounding Pr corner atoms at $\mib{R}=(\pm 1,\pm 1,\pm 1)$.  Summation over eight sites gives the form factor as
\begin{equation}
F_z(\mib{k}) = \sqrt{8}{\rm i}\cos k_x\cos k_y
\sin k_z,
\end{equation}
where $\mib{k}$ is the wave number of the conduction electron.  
Other components $F_\nu(\mib{k})$ with $\nu=x,y$ are obtained from the case of $\nu=z$
by cyclic rotation of $k_x, k_y, k_z$.
Let us introduce a set $\nu =x,y,z$ of new Wannier orbitals each of which
is created by the operator 
\begin{equation}
\psi_{\nu\sigma}^\dagger = \frac{1}{\sqrt{N}}
\sum_{\mib{k}} F_\nu (\mib{k}) c_{\sigma}^{\dag} (\mib{k}). 
\label{wannier}
\end{equation}
Then the effective hybridization through $t_u$ orbitals is given by
\begin{equation}
H_{\rm t} = \frac{1}{\sqrt{N}}
\sum_{\beta}\sum_{\nu \sigma} 
\left[ 
V_{\beta} 
\psi_{\nu\sigma}^\dagger f_{\beta \nu \sigma}
	 +\text{H.c.} 
\right],
\label{tu}	 
\end{equation}
where $\beta$ distinguishes two different sets with the $t_u$ symmetry in $4f$ orbitals such as 
$3z^3-5$ and $z(x^2-y^2)$. 
The effective hybridization parameter $V_{\beta}$ is determined by $V_{t{(\beta\alpha)}}$,  the intersite mixing between $t_u(\alpha)$ and $a_u$, and 
their energy separation.

Following the same line as derivation of the effective magnetic exchange, the second order effects of eq.(\ref{tu}) gives quadrupole coupling.
To describe the coupling we introduce the operator 
$O_{\mu\nu}^{(\gamma)}$ where $(\gamma)=(1\rightarrow 5)$ represents the quadrupole from the CEF singlet-triplet transition, 
while $(\gamma) = (5\rightarrow 5)$ comes from the intra-triplet transition.  
We shall argue later that the latter quadrupole operator is a key to understand the strange heavy-fermion behavior found in PrOs$_4$Sb$_{12}$.
The effective quadrupole coupling between $4f$ and conduction electrons is given by
\begin{equation}
H_{\rm quad} = 
\sum_\gamma 
I_{\rm quad}^{(\gamma)}\sum_{\mu\nu} O_{\mu\nu}^{(\gamma)} 
\sum_\sigma
\psi_{\nu\sigma}^\dagger \psi_{\mu\sigma }.
\label{quad}
\end{equation}
We shall evaluate these two coupling constants $I_{\rm quad}^{(\gamma)}$ 
quantitatively in separate work.

\section{Fluctuating quadrupoles}

\subsection{Triplet pair with AFQ interaction}

An AFQ order of the $\Gamma_3$-type 
involving a CEF triplet involves a subtle problem concerning the degeneracy in the ordered phase. 
The normalization of relevant operators is defined by
\begin{equation}
O_2^0 = 3^{-1/2}[3J_z^2-J(J+1)], \ \ 
O_2^2 = J_x^2-J_y^2.
\end{equation}
In the case of $\Gamma_4$ triplet, nonzero matrix elements are given by
$\langle 0|O_2^0|0\rangle =28/\sqrt{3}, \langle \pm |O_2^0|\pm \rangle =-14/\sqrt{3}$, and
$\langle \pm |O_2^0|\mp \rangle =-14$.
For identifying the degeneracy problem, which is intimately related to the 
ordering pattern in PrFe$_4$P$_{12}$, 
we take a two-site system where the triplet $\Gamma_4$ is present
at each site $A$ or $B$.
The interaction Hamiltonian is given by
\begin{eqnarray}
H_{\Gamma 3}= g_3 (O_{2,A}^{0}O_{2,B}^{0}+O_{2,A}^{2}O_{2,B}^{2}). \label{tsham1}
\end{eqnarray}
We recombine three states in the triplet as 
\begin{equation}
|z\rangle = |0\rangle, \ \ 
|x\rangle = \frac{1}{\sqrt{2}}(|+\rangle+|-\rangle ), \ \ 
|y\rangle = \frac{1}{\sqrt{2}{\rm i}}(|+\rangle-|-\rangle ).
\end{equation}
Then we take the nine-fold basis $| k\rangle_A | l\rangle_B$ for the two sites, where $k,l$ are $x$, $y$ or $z$.
By diagonalization of $H_{\Gamma 3}$
we find that the nine-fold degeneracy splits into  six-fold and three-fold degenerate states. 
With antiferro-type interaction $g_3>0$, the ground state is the six-fold multiplet with energy $-196g_3/\sqrt{3}$, and
the three-fold degenerate excited level has energy  $784g_3/3$.   
The six-fold degeneracy comes from the number of ways to choose different orbitals for $|k\rangle_A |l\rangle_B$
with $k\neq l$.
Because of the absence of off-diagonal elements in the two-site Hamiltonian, 
the mean-field theory at zero temperature becomes exact for the two-site system.
The bcc lattice can be separated into $A$ and $B$ sublattices.
Then with only the $\Gamma_3$-type intersite interaction, 
the degenerate ground state of the bcc lattice should be given exactly by the mean-field theory.

Thus a model with nearest-neighbor AFQ interaction of $\Gamma_3$ quadrupoles 
has a degeneracy with respect to different kinds of quadrupolar patterns.
In this sense, the $\Gamma_3$ antiferro-quadrupolar ordering model within the triplet state is similar to the three-state antiferromagnetic Potts model,  which possesses a macroscopic degeneracy in the ground state. 

\subsection{Interaction with conduction spins}
\begin{figure}[tb]
\centerline{
\includegraphics[width=0.9\linewidth]{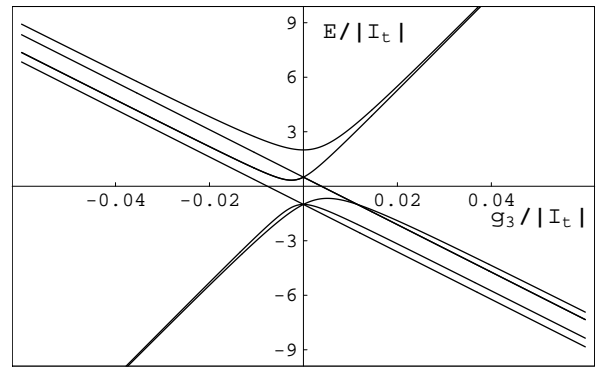}
}
\caption{Energy levels of the triplet pair each of which is coupled with a spin in the conduction level.}
\label{levels}	
\end{figure}

We now ask how the degeneracy should be broken in actual PrFe$_4$P$_{12}$.
Experimentally, distortion of Fe sublattice has been observed by X-ray and neutron diffraction.
Although the distortion is suggested to be of $O_2^0$-type \cite{iwasa3} for each Fe cube,  the diffraction results themselves do not exclude a
possibility that the actural distortion is of a breathing type as found in PrRu$_4$P$_{12}$. 
We consider a situation where conduction electrons cause quantum fluctuations among degenerate AFQ configuration.
If the macroscopic degeneracy is removed by this fluctuation mechanism, 
the cubic symmetry is kept on the average, but
a strong two-site AFQ correlation dominates the ground state.
In order to pursue the idea qualitatively, we consider a toy model
in which a conduction-electron spin is introduced at each site, and each spin has   exchange coupling with a triplet.  
Namely the local interaction $H_{\rm exc}^{(A)}$ is given by eq.(\ref{XtXs}) for site $A$, and $H_{\rm exc}^{(B)}$ for site $B$.  
Let us consider the system
\begin{equation}
H_{\Gamma3+{\rm exc}} = H_{\Gamma 3} +
H_{\rm exc}^{(A)} + H_{\rm exc}^{(B)},
\end{equation}
where we neglect the $\Gamma_1$ singlet for simplicity.
According to microscopic theoretical result \cite{Otsuki1}, we take $I_{\rm t}<0$.
In the limit $g_3=0$, we obtain the quartet ground state and the doublet excited state for each site with separation $3|I_{\rm t}|/2$.  
Thus the whole system is 16-fold degerate in the ground state.
The AFQ interaction $g_3$ breaks the degeneracy.
In the case of large $g_3/|I_{\rm t}|$, on the other hand, the lower manifold of $6\times 4=24$ states has
the following level sequence from the ground state:
(i) quartet, (ii) octet, (iii) octet, (iv) quartet.
The degeneracy of (i) comes from the spin degrees of freedom. 
Since there is no magnetic interaction between the pair, the Kramers degeneracy at each site cannot be broken.
Figure \ref{levels} shows the energy levels as a function of $g_3/|I_{\rm t}|$ taking the triplet as $\Gamma_4$.  
The asymptotes for $|g_3/I_{\rm t}|\gg 1$ are characterized by the ferro-orbital state with the slope $784/3$ as obtained earlier, 
and the antiferro-orbital state with the slope $-196/\sqrt{3}$.   The ratio of the slopes is $-4/\sqrt{3}\sim -2.3$.
 
\begin{figure}[b]
\centerline{
\includegraphics[width=0.9\linewidth]{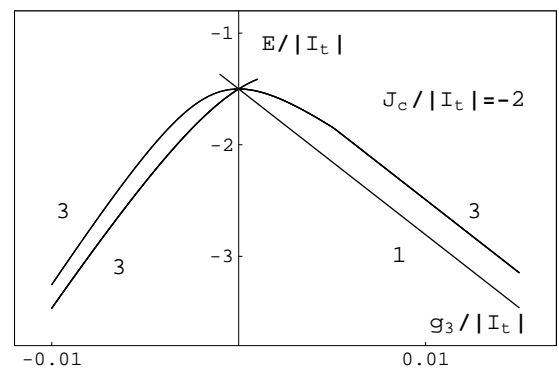}
}
\caption{The lowest two energy levels of the pair with inclusion of intersite coupling $J_c$ of conduction spins.  The numbers 1, 3 show the degeneracy of each level.}
\label{levels2}	
\end{figure}

Now we include an intersite magnetic interaction of the form
\begin{equation}
H_{\rm c2} = J_{\rm c} \mib{s}_c^{(A)} \cdot \mib{s}_c^{(B)}.
\end{equation}
The ground-state quartet obtained above
is split by this interaction.
A singlet ground state emerges in the case of $J_{\rm c}<0$, and a triplet ground state otherwise.
Figure \ref{levels2} shows the lowest two levels for both signs of $g_3/I_{\rm t}$.

The wave function of the ground state for the system 
$H_{\rm pair}=H_{\Gamma3+{\rm exc}}+
H_{\rm c2}$ 
can be derived explicitly.    We introduce notations:
\begin{eqnarray}
|x,y\rangle_{\pm} &=& 2^{-1/2}\left[ 
|x\rangle_A |y\rangle_B \pm |y\rangle_A |x\rangle _B  \right], \\
|s_z;0\rangle &=& 2^{-1/2}\left[ 
|\uparrow\rangle_A |\downarrow\rangle_B + |\downarrow\rangle_A |\uparrow\rangle _B  \right], 
\end{eqnarray}
for a pair of CEF states and a pair of conduction spins, respectively.
By cyclic rotation, we define similar states such as $|yz\rangle_\pm$ and 
$|s_x;0\rangle$.  The latter is given by
\begin{equation}
|s_x;0\rangle = 2^{-1/2}\left[ 
|\uparrow\rangle_A |\uparrow\rangle_B + |\downarrow\rangle_A |\downarrow\rangle _B  \right],
\end{equation}
which has spin 1 with component $s_x=0$.
Then the ground state $\psi_0$ of $H_{\rm pair}$ 
with $g_3>0, \ \ I_{\rm t}<0$ and $J_{\rm c}<0$
is given by
\begin{equation}
\psi_0 = 
3^{-1/2}\left[ 
|xy\rangle_+ |s_z;0\rangle +
|yz\rangle_+ |s_x;0\rangle +
|zx\rangle_+ |s_y;0\rangle  \right], 
\end{equation}
which does not depend on the interaction parameters because of its high symmetry.
Although the pair has a complete AFQ correlation,  there is no finite static quadrupole at each site.  As a result, the cubic symmetry at each site is preserved on the average.

\subsection{Staggered moments induced by magnetic field}

The dominance of quadrupolar moments in the ordered phase in PrFe$_4$P$_{12}$ is indicated by 
the results of polarized neutron scattering\cite{hao} 
where the induced AFM moments is about 2 times larger for the field direction $(001)$ than for $(110)$. 
However, there is no component of staggered magnetization perpendicular to the magnetic field.  If there is a static quadrupole order, one may naturally expect some anisotropy leading to the perpendicular component.   
The absence is not due to cancellation by the mixture of quadrupole domains, since NMR has observed a sharp induced staggered field \cite{kikuchi}. 

If we assume the fluctuating state with no static AFQ order as described by the pair, there is no perpendicular component in the staggered magnetiztion.
In our scenario of fluctuationg AFQ order, the static order parameter in PrFe$_4$P$_{12}$ is a monopole, hexadecapole (rank 4 tensor) and/or a hexacontatetrapole (rank 6 tensor).  Both tensors can form a scalar in the poing group $T_h$ as given by $O_4, O_6^{\rm c}$ and $O_6^{\rm t}$ in eq.(\ref{V_CEF}).  Similar multipoles have been postulated as order parameters in PrRu$_4$P$_{12}$.  
An important difference, however, is the large N\'{e}el-like anomaly in the magnetic susceptibility in PrFe$_4$P$_{12}$, and its absence in PrRu$_4$P$_{12}$.
We suspect the AFQ interaction is insignificant in PrRu$_4$P$_{12}$.

\subsection{Uniaxial pressure effect}
\begin{figure}[tb]
\centerline{
\includegraphics[width=0.8\linewidth, angle=270]{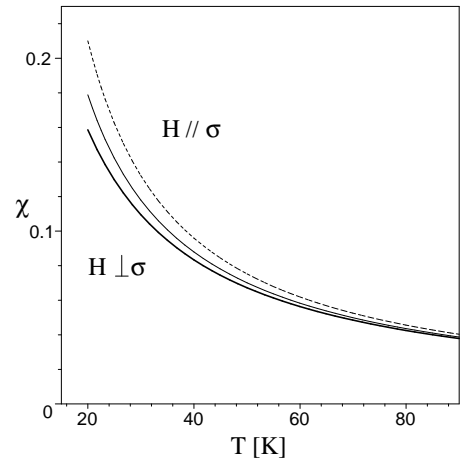}
}
\caption{
Magnetic susceptibility of an isolated pseudo-quartet under uniaxial pressure $\sigma = 3$kbar.
The parameters are taken as $g_3=0.3$K and
$\Delta=3$K. The middle line shows the susceptibility without
uniaxial stress.  The unit of $\chi$ is $(g_J\mu_B)^2/T_0$ with $g_J=4/5$ and 
$T_0=1$ K.}
\label{chi-stress}	
\end{figure}

The quadrupole fluctuation is much affected by uniaxial stress, which serves as valuable probe to identify the nature of electronic states.
According to ref.\citen{Matsuda}, the susceptibility is much enhanced by uniaxial pressure along the magnetic field, while it is insensitive perpendicular to the field.  The large anisotropy is common to both ordered and disordered phases.
In order to understand the origin of such anisotropy, we have checked how the single-site triplet is affected by uniaxial stress.  
Provided the uniaxial stress along the $z$-axis favors oblate charge distribution of $4f$ states, the CEF triplet splits into 
the ground-state doublet and a singlet.  Since the doublet $|\pm\rangle$ has a moment along the $z$-axis, the susceptibility $\chi_z$ is enhanced.  On the other hand, $\chi_x$ and $\chi_y$ come from transition from the doublet and the singlet, which is reduced by the stress.
Figure \ref{chi-stress} shows an example of model calculation for a $\Gamma_1-\Gamma_4$ pseudo-quartet.
Here the magnetoelastic coupling is taken to be $g_{\Gamma3}=100$K, and the elatic constant 
$C_{11}-C_{12}=35$GPa \cite{nakanishi}.
It is seen that the decrease of the suscepibility for the perpendicular field is half of the increase along the field.  This relation comes from the invariance of the average susceptibility in the present model.  
Although the tendency of anisotropy is consistent with experimental observation, the quantitative discrepancy is serious.  It seems necessary to include fluctuation effects even in the disordered phase.

\section{X-ray detection of hexadecapole order}
\begin{figure}[bt]
\centerline{
\includegraphics[width=0.9\linewidth]{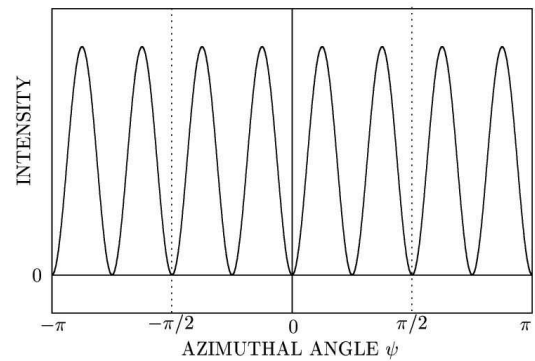}
}
\caption{
Predicted azimuthal angle dependence from the hexadecapole order in 
PrRu$_4$P$_{12}$ and PrFe$_4$P$_{12}$.  Assumed are the (001) surface and the $\sigma$-$\pi'$ scattering.
}
\label{azimuthal}	
\end{figure}

We discuss how to probe possible hexadecapole order transition in 
 PrFe$_4$P$_{12}$ and  PrRu$_4$P$_{12}$.
The most direct observation should be azimuthal angle scan in resonant X-ray scattering.  In the quadrupole scattering process called E2, the staggered order of hexadecapoles contributes to the superlattice scattering.
On the other hand, the hexacontatetrapole cannot be probed by the E2 scattering because of its higher rank $(=6)$.
The combination $O_4^0+5O_4^4$ constitutes a scalar in the cubic symmetry, which has a staggered component in the hexadecapole order.  
We have shown \cite{kusunose} that the amplitude should depend on the azimuthal angle $\psi$ as 
$\cos 4\psi$ in the $\sigma$-$\sigma'$ channel,
and as $\sin 4\psi$ in the $\sigma$-$\pi'$ channel.
In the $\sigma$-$\pi'$ channel, the polarization of the light is rotated by 90$^\circ$  from the incident beam.  
These rapid oscillations of the amplitude are not shared by amplitude for other multipoles.  
In particular, there is no contribution from $O_4^0$ in the $\sigma$-$\pi'$ channel.
The resulting intensity has the eight-fold pattern ($\propto 1-\cos 8\psi$) with a minimum at  $\psi =0$.   The $\sigma$-$\sigma'$ channel will show a four-fold pattern by combination with $O_4^0$. 
The intensity pattern is illustated in Fig.\ref{azimuthal}.  
In this way resonant X-ray scattering using the E2 transition can identify the hexadecapole order.    
On the other hand, the non-resonant scattering also has contribution from hexadecapoles with angular momentum $L=4$, which can be identified by the form factor analysis.
Although the monopole contribution should dominate the intensity, 
one may hope to identify the contribution from $L=4$  since the quadrupole component $L=2$ is absent.  
Identification of the $L=4$ component provides  alternative detection of the hexadecapole order.

\section{Possible origin of heavy mass in PrOs$_4$Sb$_{12}$} 

\begin{figure}[tb]
\centerline{
\includegraphics[width=0.9\linewidth]{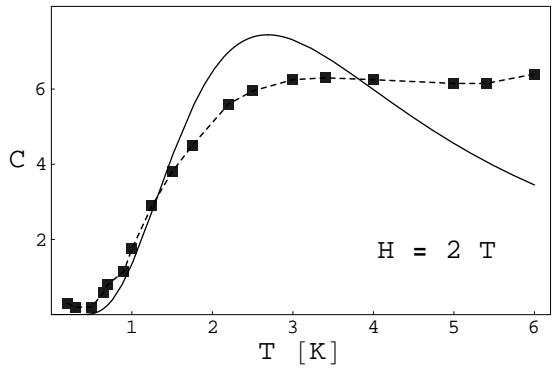}
}
\caption{Comparison of the Schottky-type model for the specific heat with experimental results in ref.\citen{aoki-jpsj}. The unit of $C$ is J/(mol$\cdot$K).
}
\label{Schottky}	
\end{figure}

The specific heat of PrOs$_4$Sb$_{12}$ has a conspicuous anomaly of the Schottky type,  which has been ascribed to the singlet-triplet CEF splitting.  The superconducting transition occurs near the peak of this anomaly with a large jump ($\sim 1$J/mol$\cdot$K), which indicates a huge mass of the Cooper pair.
If the superconductivity is destroyed by magnetic field, however, the heavy itinerant electrons seem to disappear at lower temperatures. 
Namely, the specific heat does not have the $\gamma T$ term with large $\gamma$.
Figure \ref{Schottky} compares the specific heat of the singlet-triplet CEF model in magnetic field with experimental results \cite{aoki-jpsj}.  
The field of $H=2$T already masks the structure from the superconducting transition. 
It is seen that the discrepancy from the CEF model becomes larger for $T >1.5$K, and the discrepancy above 4 K grows linearly with $T$, which gives the proportionality constant $\gamma \sim 0.9 $J/(mol$\cdot$K$^2$).
In the low temperature end, on the other hand, the Schottky model gives reasonable account of the observed result.   
We therefore conclude that a fraction of the entropy above $\sim$1K is transferred to the temperature range higher than the maximum of $C(T)$.   
On the basis of observation above, we speculate that a quadrupole Kondo effect is active in this system where intra-triplet quadrupole excitations scatter conduction electrons.  The mass enhancement from the intra-triplet excitation channel should naturally decrease if the thermal population of the triplet becomes small.   
On the other hand, the singlet-triplet quadrupole channel should saturate below the temperature corresponding to the CEF splitting.
The lowest order effect of this latter channel has been discussed by Fulde and Jensen \cite{fulde}.  
Since the $\gamma T$  term disappears at low temperature in
Fig.\ref{Schottky},  it seems that the intra-triplet channel is the dominant source of the heavy-fermion behavior.  Microscopic study of $I_{\rm quad}^{(\gamma)}$  in eq.(\ref{quad}) will shed further light on the validity of this hypothesis.

In contrast with the ordinary Kondo-type mechanism, the quadrupole scattering of conduction electrons with the $a_u$ symmetry requires a coherence of Pr atoms of the order of 10
as shown by eq.(\ref{wannier}).  Hence it should be more sensitive to disorder such as La doping \cite{rotundu} than the ordinary Kondo effect.  
It should be a difficult but interesting problem how the heavy Cooper pairs survive or disappear in PrOs$_4$Sb$_{12}$ at very low temperature where thermal population of the triplet becomes negligible.

\section{Summary and Outlook}

In this paper we have proposed some ideas for understanding the intriguing properties of Pr skutterudites.  The key to the understanding is the nature of the triplet CEF states and the singlet-triplet splitting which vary among  PrFe$_4$P$_{12}$, PrRu$_4$P$_{12}$ and PrOs$_4$Sb$_{12}$.   The difference of CEF states explains the presence of the strong Kondo effect in PrFe$_4$P$_{12}$, and the sharp CEF excitations observed in PrRu$_4$P$_{12}$ and PrOs$_4$Sb$_{12}$. 
Recently Sm skutterudites have also been actively investigated.   In particular, 
SmOs$_4$Sb$_{12}$ has a large $\gamma T$ term in the specific heat, which is insensitive to magnetic field.  We suspect that quadrupolar Kondo effect may be the origin of the heavy mass.   In contrast with the case of PrOs$_4$Sb$_{12}$, the quadrupole active states are in the ground CEF level.   These ideas can be corroborated only by microscopic calculation based on a realisitic model.   We plan to test the above ideas in the near future.

The authors thank O. Sakai, K. Iwasa and M. Kohgi for informative discussion.
This work was supported partly by 
Grants-in-Aid for Scientific Research on Priority Area ``Skutterudite" , and 
for Scientific Research (B)15340105 
of the Ministry of Education, Culture, Sports, Science and Technology, Japan.

\end{document}